\documentstyle[aps,twocolumn]{revtex}

\begin{document}
\title{THERMALLY ASSISTED TUNNELING: AN ALTERNATIVE MODEL\ FOR THE
THERMOLUMINESCENCE PROCESS IN CALCITE}
\author{\medskip J. F. de Lima$^{\dagger }$, M. E. G. Valerio$^{\dagger }$ and E.
Okuno$^{\ddagger }$}
\address{\medskip $^{\dagger }$Depto de F\'{i}sica, Universidade Federal de Sergipe,\\
49.100-000, S. Crist\'{o}v\~{a}o-SE, Brazil, \\
e-mail: lima@ufs.br or mvalerio@ufs.br\\
$^{\ddagger }$Instituto de F\'{i}sica, Universidade de S\~{a}o Paulo, CP
66.318, \\
05315-970, S. Paulo, Brazil\bigskip }
\date{\today}
\maketitle

\begin{abstract}
{\bf Abstract:} {\it In this work we investigate the effect of gamma and
ultraviolet radiations on the thermoluminescence of Brazilian calcite. The
irradiated samples presented three TL peaks at 150, 245 and 320}$^o${\it C
(at a linear heating rate of 1.8}$^o${\it C/s), with a main emission band
centered at 615nm due to the }$^4${\it G }$\rightarrow ${\it \ }$^6${\it S
transition of the Mn}$^{2+}${\it . The irradiated samples presented, besides
the Mn}$^{2+}${\it \ signals, three lines related related to carbonate
groups. The thermal treatment and the irradiation effects on the ESR signal
lead to the conclusion that (CO}$_3${\it )}$^{3-},${\it \ stabilized in two
different symmetries, and the (CO}$_2${\it )}$^{-}${\it \ ions are the
electron trapping centers. It was found that the TL peaks follows a t}$^{-1}$%
{\it \ decay as a function of the UV illumination time. An alternative model
for the TL emission of calcite is discussed, considering that the
recombination of charges is processed via a thermally assisted tunnelling
mechanism.}
\end{abstract}


\section{Introduction}

Calcite is a mineral found in many geological formations. Its
thermoluminescence (TL) has been studied, mainly due to its application in
geological and fossil dating\cite{1,2,3}. Calcite can also be used for
ionizing and ultraviolet radiation dosimetry.

Investigations on calcite have shown different TL characteristics depending
on the impurity content and on the genesis of the sample. It was also
observed on artificially grown CaCO$_3$ crystals that the type and
concentration of impurities modify both TL sensitivity and glow curve shape.

Important contributions have been made to the understanding of the TL
mechanism of calcite since the pioneering works of Kolbe and Smakula\cite{4}
and Medlin\cite{5} in the sixties. Studying the influence of manganese
concentration, and by comparison of the emission spectrum of calcite with
the emission of the others lattices including manganese, Medlin found that
the emission band centered in 615nm, corresponding to a transition from the
first excited to the fundamental state of Mn$^{2+}$ ion occupying a Ca$^{2+}$
site. Recent works\cite{6,7} showed that the shape of the emission spectrum
changes from a broad band without structure to a well-resolved narrow band
when the temperature of the TL peak increases and when the concentration of
the Mn decrease. They also concluded that the orange emission is related to
Mn impurity. Calderon at al\cite{8} suggested that the holes are trapped by
the impurities and electrons are trapped at dislocations forming (CO$_3$)$%
^{3-}$ . The irradiation with UV light was found to modify the intensity of
the TL peaks suggesting that the trapped carriers can be phototransfered
from deep to shallow traps \cite{9}. Concerning the kinetics of the
detrapping process, it was found that the isothermal decay follow a t$^{-1}$
law which could not be accounted for a single trap model. Many workers
explained this results assuming that the process are due to a distribution
in the activation energies\cite{13,14,15}. On the other hand, Visocekas et al%
\cite{10} suggested that, for the peaks observed on their samples below room
temperature, this behavior could be due to a tunnelling process.

In the present work we investigate the effect of gamma and ultraviolet
radiations on the thermoluminescence of Brazilian calcite from Miranda, MS.
Based on a correlation between TL and electron spin resonance (ESR) and on
the photoinduced TL decay, we suggest an explanation for the appearance of
the three peaks between room temperature and 450$^o$C, and propose an
alternative model for the thermoluminescence process.

\section{Experimental}

Brazilian calcite from Miranda, MS used in this work was powdered and sieved
and the grains with diameters between 74$\mu $m and 174$\mu $m were
selected. Chemical analysis by atomic absorption and X-ray fluorescent
spectroscopy techniques detected the following impurities: Si(%
\mbox{$<$}%
0.1\%), Al(%
\mbox{$<$}%
0.1\%), Fe(%
\mbox{$<$}%
0.05\%), Ti(%
\mbox{$<$}%
0.05\%), Mg(%
\mbox{$<$}%
0.23\%), Na(%
\mbox{$<$}%
0.03\%) and K(%
\mbox{$<$}%
0.05\%). Electron spin resonance measurements revealed the presence of
manganese. Samples were exposed to $^{60}$Co gamma rays from 50Gy to 500Gy.
Ultraviolet irradiation was performed at a distance of 0.5 m from a high
pressure 400W mercury lamp. A home made TL reader was used to heat 4 mg
powder samples from 30$^o$C to 400$^o$C in a nitrogen atmosphere, at a
heating rate between 1$^o$C/s to 3$^o$C/s. We performed a series of
monochromatic TL measurements with wavelength varying from 450 to 750nm in
steps of 10nm. Data were stored in a microcomputer and the 3D isometric
plots were constructed via software after correcting each of monochromatic
curves to the wavelength response of the system.

ESR experiments were performed with a BRUKER ER 200D equipment, also on
powdered samples at room temperature, in the X-band and 1.0 mW microwave
power (below saturation). The peak to peak height of the first derivative of
the signal was considered to be proportional to the number of paramagnetic
centres.

\section{Results}

The glow curve of untreated sample (without any treatment in laboratory)
consists of two peaks at 245 and 320$^o$C when heated at a rate of 1.8$^o$%
C/s. After a laboratory dose of gamma rays, the intensity of these peaks
increases and another peak, around 150$^o$C, appears. Hence for simplicity
these peaks are labeled as peaks A, B, and C as their temperatures increase.

Figure 1 shows the isometric plot and the contour plot of the TL spectra of
the irradiated samples. The three main peaks present a broad emission band
centered at 615 nm. Besides this band, peaks A and B present a very weak
emission band at 480nm, not usually reported. Comparing the TL spectra
around 615nm at the peak temperatures we can see that peak C displays a
fine-structure with at least 4 emission lines in this region while peak A
are mainly composed by the main emission at 615nm. These results agree with
previous results\cite{6,7}. Treated samples (with a heat treatment at 400$^o$%
C for 1 hour followed by a gamma irradiation) show a similar glow curve,
without changes in the peak positions.

The growth and the saturation of the peak heights with the exposure to gamma
rays are shown in figure 2. The effect of post-irradiated thermal treatment
at different temperatures shows that the decrease of TL intensity is
accompanied by the dislocation of the TL peak maxima to higher temperatures
indicating that the TL processes can not be accounted for a first order
kinetic. Figure 3 shows the effect of 10 min post-irradiation annealing at
different temperatures. We can see that as the temperature of
post-irradiation thermal treatment increases, the intensity of the TL peaks
decrease. Peak A vanish first followed by peak B and finally by peak C.

The effect of ultraviolet (UV) irradiation on the thermoluminescent
properties was studied using both untreated (without any treatment in the
laboratory) and treated samples (with a heat treatment at 400$^o$C for 1
hour followed by gamma irradiation). When treated samples are irradiated
with UV all the three peaks decrease and the peak maxima shift to higher
temperatures (Figure 4), a similar shifts is produced by thermal bleaching.
To check if any isothermal reduction of the peak was happening during the UV
illumination, the temperature of the samples was monitored. We found that
during the UV illumination the temperature is around 30$^o$C. Measurements
of the isothermal decay of the TL peaks revealed that, considering the time
intervals used in these experiments, the isothermal decay at this
temperature can be neglected. Figure 5 shows the variation of the inverse of
TL intensity as a function of the illumination time ''{\bf t ''} where we
can see that TL intensity of the three peaks are proportional to t$^{-1}$.

In figure 6 we show the result of the intensities of the ESR signals as a
function of the radiation dose. The ESR measurements were performed at room
temperature and it is possible to see that increasing the gamma ray doses,
the well known signals of the Mn$^{2+}$ ion present in the samples decrease.
At the same time three other signals labelled 2, 3 and 5, with g$_2$=2.0036,
g$_3$=2.0032, and g$_5$=2.0006 typical of free radicals, increase. The
effect of the post-irradiation thermal treatment at different temperatures
during 10 minutes on the labeled ESR signals are shown in figure 7. The
increase of temperature of post irradiation heat treatments causes the
increase of the Mn$^{2+}$ signal and decrease firstly of the signal 3, at
the 75-150$^o$C range, followed by the signal 2, at 200-350$^o$C, and
finally the signal 5, starting at 280$^o$C (Figure 7). Thermal treatment at
400$^o$C during 1 hour eliminate all the 2, 3 and 5 signals produced by the
irradiation.

\section{Discussion}

At present, the plain understanding of thermoluminescent emission requires
the identification of the luminescence and the trapping centers and the
indication of the charge carrier transference and recombination processes.
The results presented on this work allow us to confirm the explanation
presented by other researches and point out an alternative model to the
charge carrier transference.

From figure 1 we can see that the three peaks have the same emission band
centered on 615nm. This result combined with the result that the ESR signal
related to Mn$^{2+}$ decrease with the irradiation dose and is recovered if
the sample is heated, (see figures 6 and 7), clearly establish that the Mn
is the hole trap formed upon irradiating the crystal and that during the
heating it is both the recombination and the luminescence center. These
results agree with the results found by other authors\cite{6,7} that pointed
out that the emission band centered in 615nm is due to the $^4$G $%
\rightarrow $ $^6$S transition of the Mn$^{2+}$ ion.

ESR signals 3 (g$_3$ = 2.0036), 2 (g$_2$ = 2.0032) have been associated\cite
{11} to the (CO$_3$)$^{3-}$ while signal 5 (g$_5$ = 2.0006) is associated
with (CO$_2$)$^{-}$. In our experiments signals 3 and 2 bleach at different
temperature ranges suggesting two different types of (CO$_3$)$^{3-}$ on the
crystal lattice. The decay of the free radical ESR signals with increasing
heat treatment temperature points out to an unambiguous correlation between
the A, B and C TL peaks and 3, 2 and 5 ESR signals. From this result,
combined with the growth of these signal as a function of the radiation dose
described in figure 6, we can conclude that the (CO$_3$)$^{3-}$ and the (CO$%
_2$)$^{-}$ are the electron trapping centers.

The results presented by Medlin\cite{13,14} showed that the isothermal decay
of the phosphorescence of calcite crystals follows a t$^{-1}$ law. Medlin%
\cite{13} and Pagonis et al\cite{15} attributed this behavior to a
recombination processes with a distribution in the energies of the electron
traps.

The T$_m$xT$_{stop}$ method, initially proposed by McKeever\cite{3} as a
tool for distinguishing among different kinetics and mechanisms, was
performed in natural calcite by Lima et al\cite{16}. Their results clearly
indicated that the distribution in activation energies can not account for
the TL peaks in calcite.

In the present work UV light was used to release the trapped electrons and
generates the electron-hole recombination process (figures 4 and 5). The
results obtained from these measurements show that the decay of the
gamma-ray induced TL peaks due to exposure to UV light is inversely
proportional to the illumination time, as we can see from figure 5.

The detrapping rate, in this experiment, is proportional to the intensity of
the UV light but the mechanism of the charge recombination and the light
emission should be the same if we employed thermal activation. In the
former, a distribution of traps can not explain the t$^{-1}$ law also
observed.

Taking into account all these results we are proposing that the charge
transfer mechanism is processed via a thermally assisted tunnelling from the
carbonate groups, as described earlier, and the Mn$^{3+}$. This process can
explain the decay of the TL peak intensity with the UV illumination time,
found in the present work, the isothermal decay of the phosphorescence in
calcite, found by Medlin and Pagonis et al., and the T$_m$xT$_{stop}$
results, found by Lima et al. This proposal is also in agreement with the
observation that the TL peaks in calcite are not accompanied by any
thermally stimulated conductivity peaks, as it was showed by Medlin\cite{17}%
, and that the emission of light during the recombination process is
generated in a ''localized-type '' mechanism\cite{18}.

If the recombination of charges between the electron and the hole traps are
via a tunnelling process, both centers should be close enough to enable the
tunnelling process to take place. Hence, we suggest that during the
irradiation of the samples the (CO$_3$)$^{3-}$ and (CO$_2$)$^{-}$ centers
are formed in the vicinity of a Mn$^{2+}$ substituting for a Ca$^{2+}$ of
the calcite matrix. The two different (CO$_3$)$^{3-}$ observed in the ESR
measurements can be attributed to carbonate groups at nearest-neighbor and
at next-nearest-neighbor positions of the Mn dopant.

\section{Conclusion}

In conclusion, the results observed in the present work clearly establish
the role of the (CO$_3$)$^{3-}$ and (CO$_2$)$^{-}$ ions as the main trapping
centres and the role of the Mn$^{2+}$ as the recombination and the
luminescence centre in the TL processes in calcite. The observed TL peaks
can be explained as follows: during irradiation, Mn$^{2+}$ ions change to Mn$%
^{3+}$, losing one electron that is captured in a (CO$_3$)$^{2-}$, forming
the (CO$_3$)$^{3-}$ ion or the (CO$_2$)$^{-}$ ion. The (CO$_3$)$^{3-}$ can
be formed either in a nn or in a nnn position to the Mn$^{3+}$ dopant
generating two different (CO$_3$)$^{3-}$ ions. Upon heating, the electron
recombines with the hole left in the Mn$^{3+}$, after a thermally assisted
tunnelling from a trap close to the manganese. The resulting Mn$^{2+}$ is
formed in an excited state, whose decay is responsible for the emitted light
during the TL process.

\pagebreak

\begin{figure}[tbp]
\caption{Isometric(a) and contour (b) plots of thermoluminescent spectra of
a gamma-ray irradiated Brazilian calcite, recorded at 1.8 oC/s.}
\label{Figure1}
\end{figure}

\begin{figure}[tbp]
\caption{TL peak heights of Brazilian calcite as a function of exposure to
gamma-rays.}
\label{Figure2}
\end{figure}

\begin{figure}[tbp]
\caption{Decay of TL intensity with annealing temperature during 10 min of
calcite previously irradiated with a laboratory dose of gamma-rays. }
\label{Figure3}
\end{figure}

\begin{figure}[tbp]
\caption{Glow curve of calcite previously exposed to gamma-rays and
illuminated with UV during 5 and 10 min. A glow curve without UV
illumination is also shown.}
\label{Figure4}
\end{figure}

\begin{figure}[tbp]
\caption{Reciprocal of the TL peak heights of thermally treated calcite as a
function of UV exposure time.}
\label{Figure5}
\end{figure}

\begin{figure}[tbp]
\caption{Intensity of ESR signals of a Brazilian calcite sample as a
function of exposure to gamma-rays.}
\label{Figure6}
\end{figure}

\begin{figure}[tbp]
\caption{Effect of the annealing temperature for 10 min in the ESR signals
of calcite irradiated with gamma-rays.}
\label{Figure7}
\end{figure}

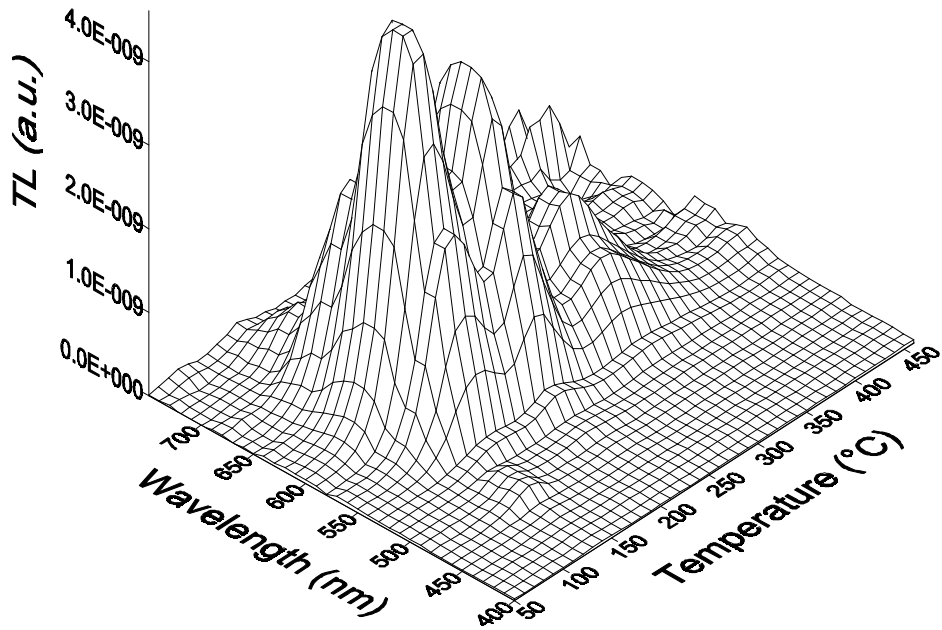
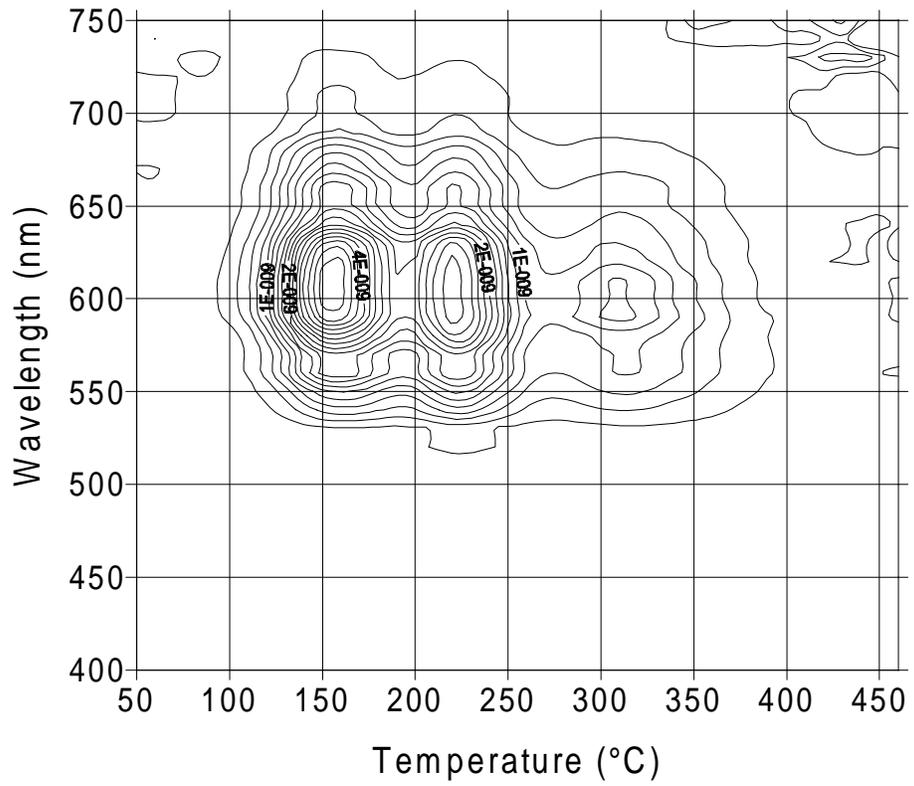
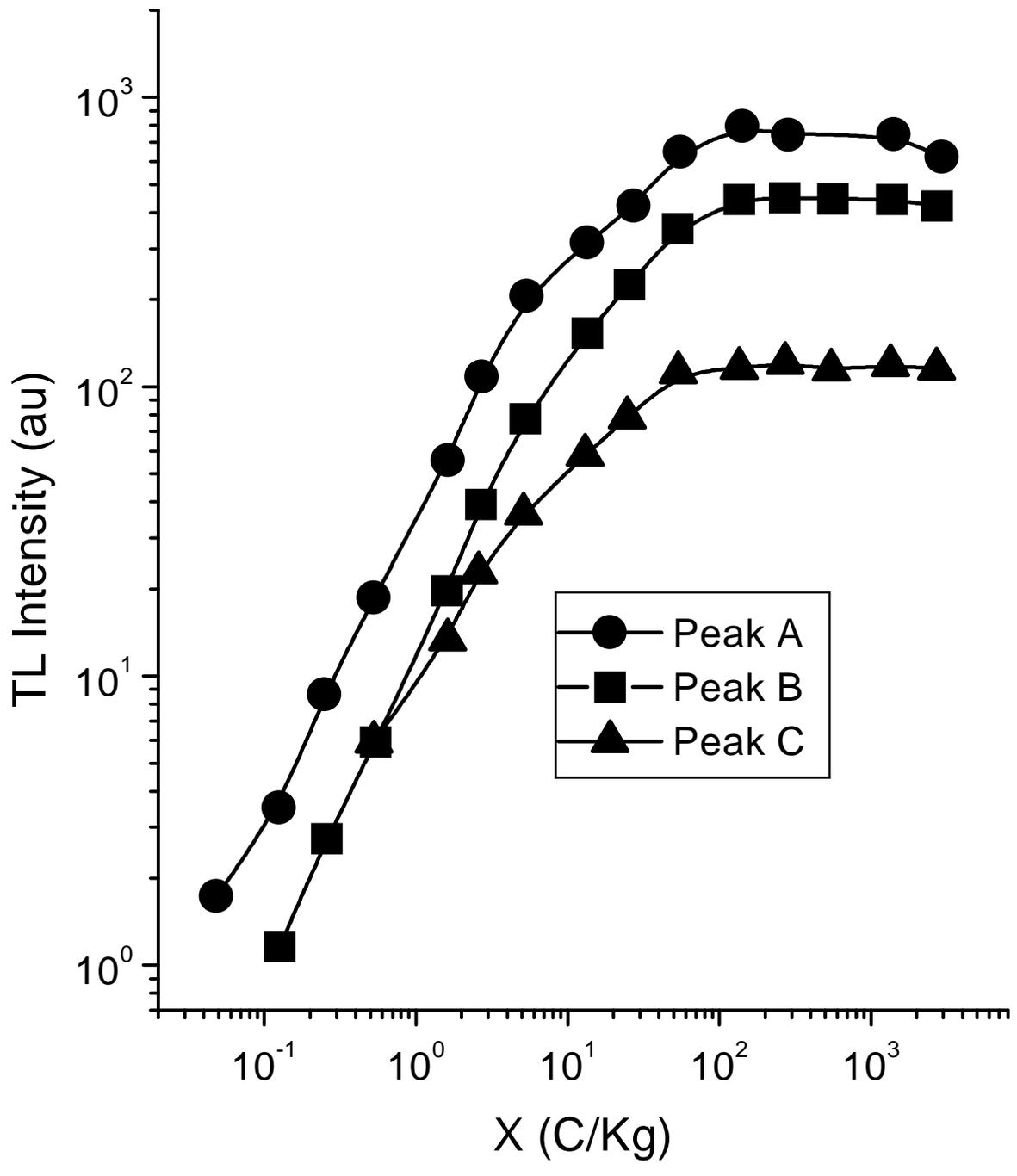
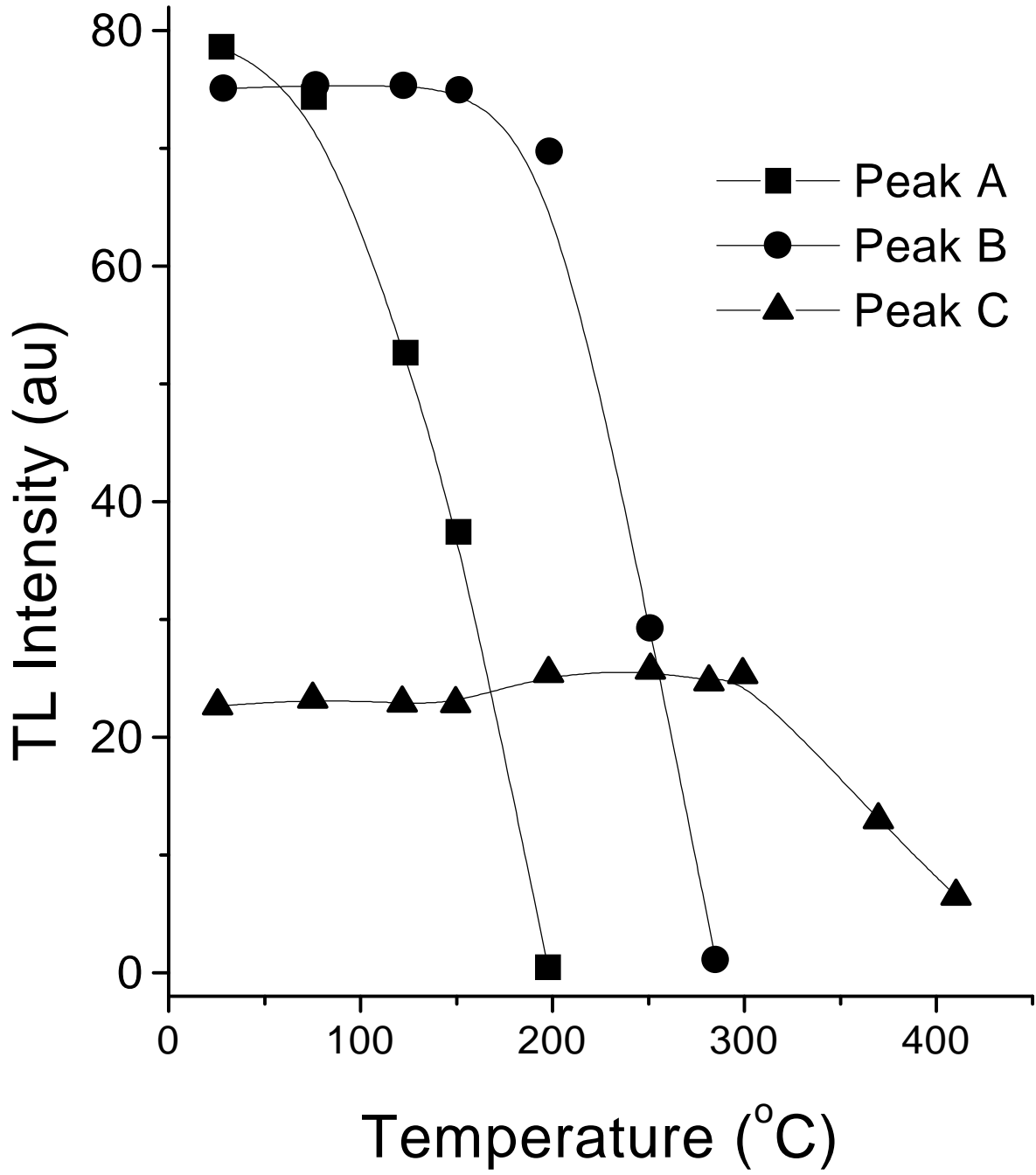
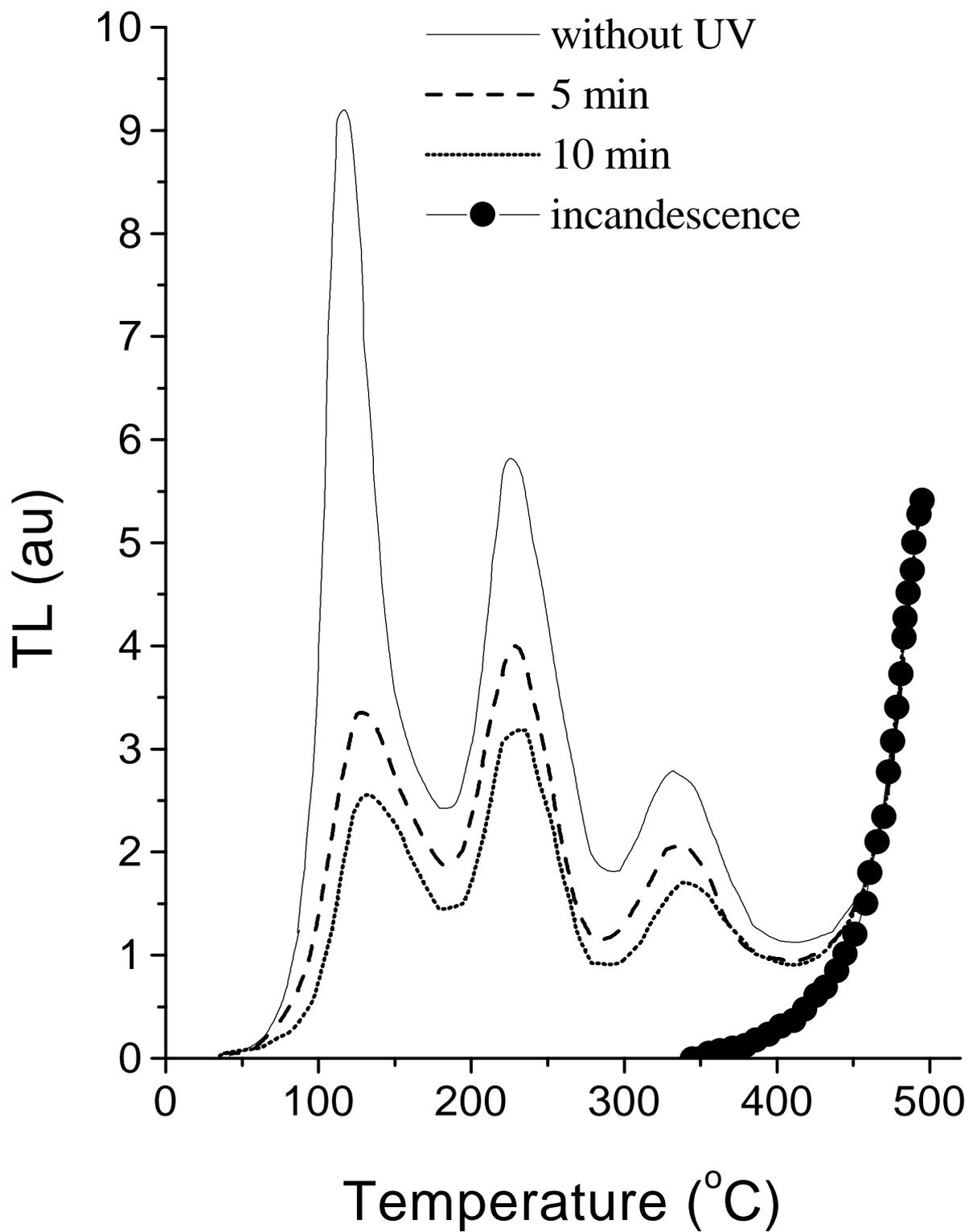
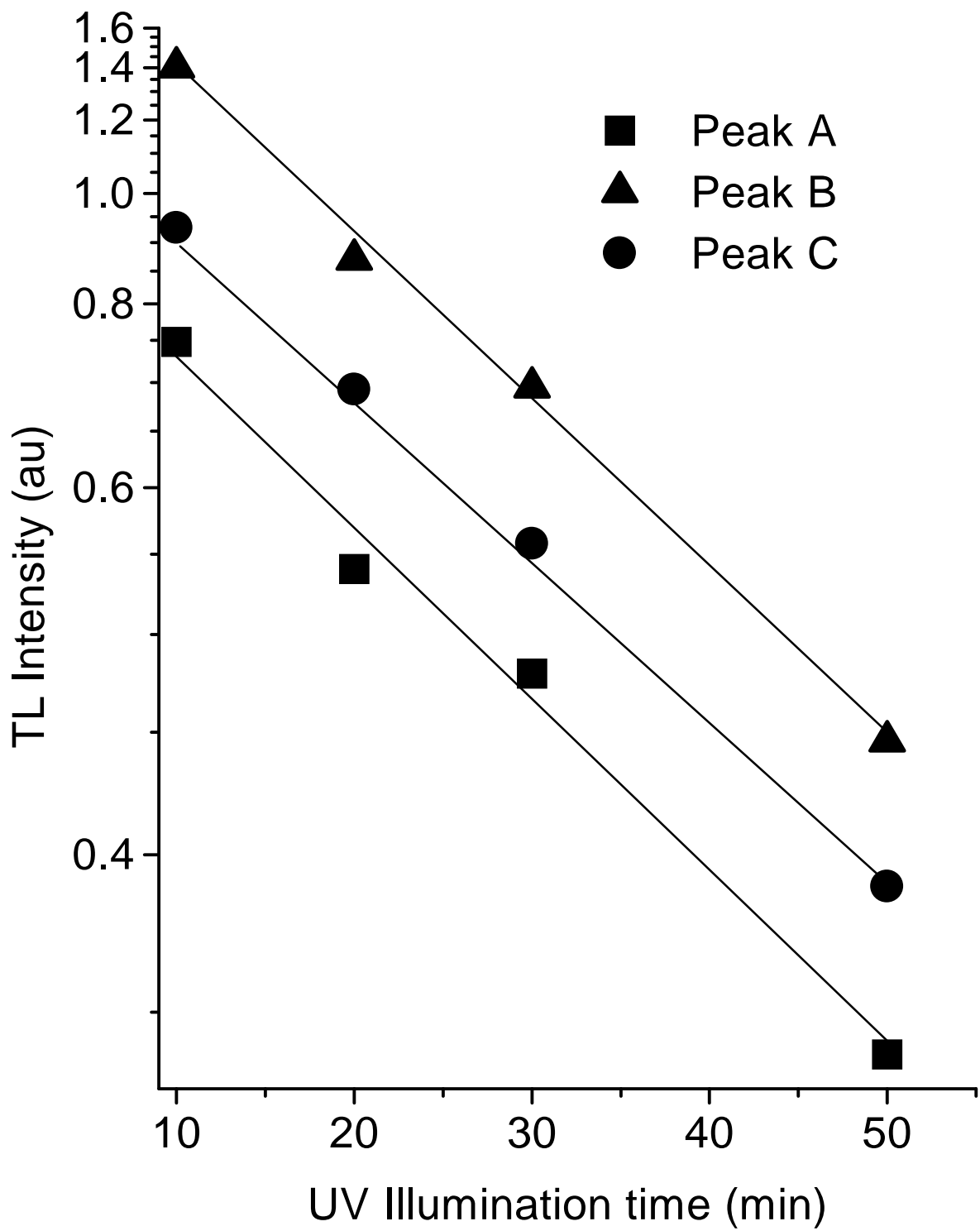
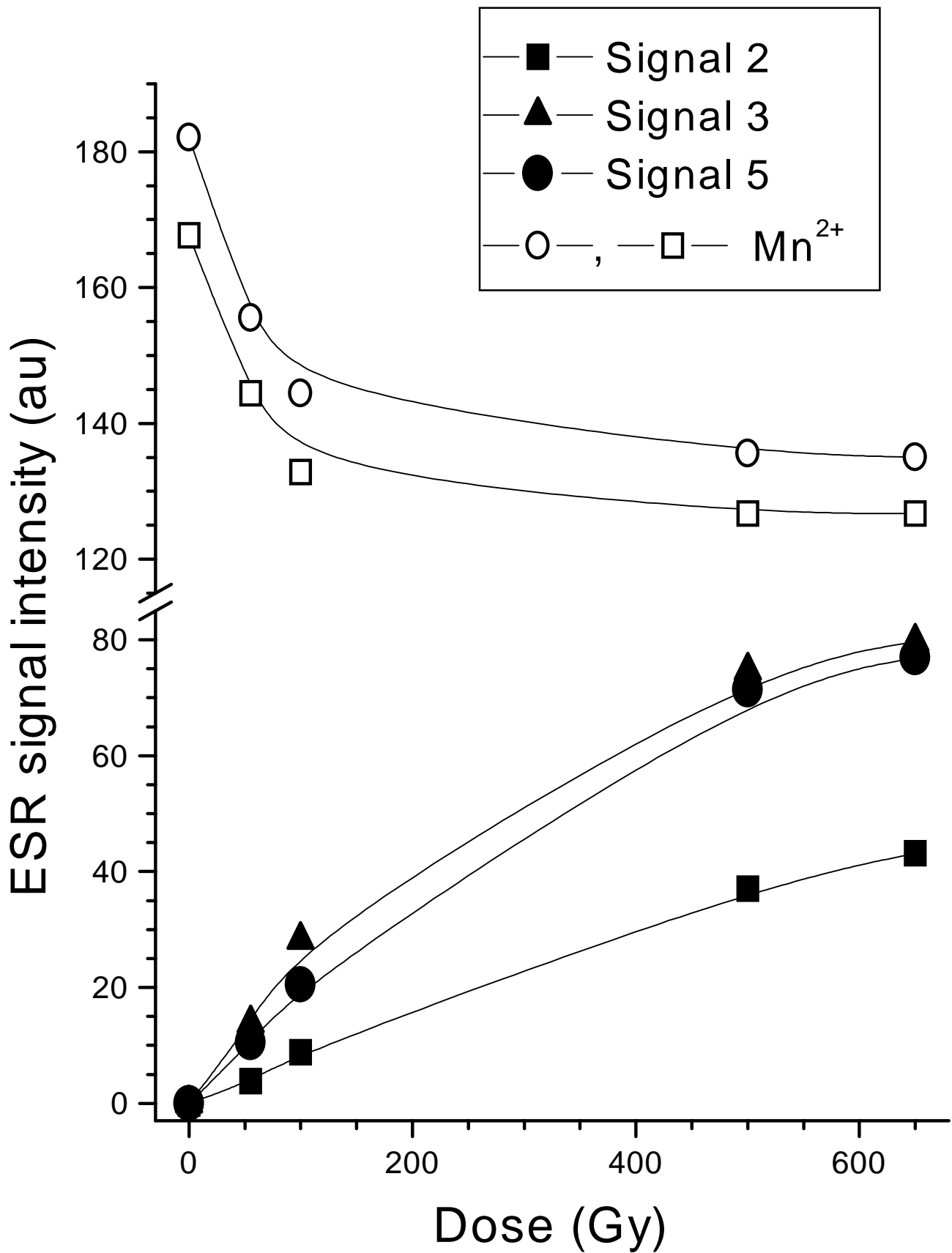
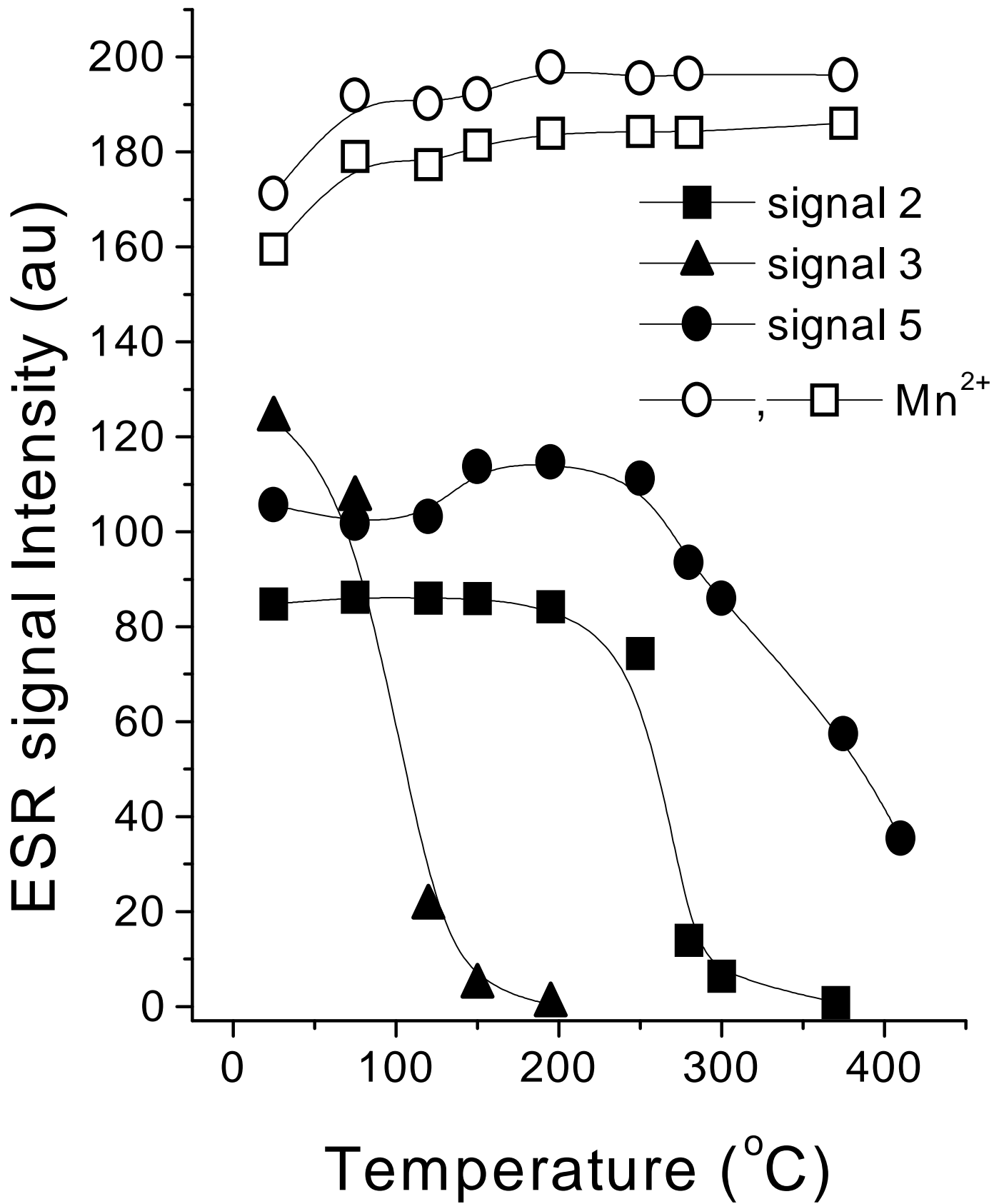
\end{document}